\documentclass[proceedings, preprint]{rmaa}

\usepackage{paralist}

\usepackage{psfrag,color}

\newcommand{\yencms}{https://bit.ly/3tmXhGo}

\newcommand{\yescms}{https://bit.ly/3FfIbVi}

\newcommand{\cm}{{\sc CosMonic}}
\newcommand{\ac}{{\sc Astroaccesible}}

\newcommand{\msun}{M$_{\odot}$}
  
\SetYear{2022}
\SetConfTitle{II Workshop on Astronomy Beyond the Common Senses for Accessibility and Inclusion}

\title{Painting graphs with sounds: CosMonic sonification project} 

\author{
  R. Garc\'ia-Benito\altaffilmark{1} 
  and
  E. P\'erez-Montero\altaffilmark{2}
  }

\altaffiltext{1}{Instituto de Astrof\'isica de Andaluc\'ia (IAA), CSIC,
  Apartado de correos 3004, E-18080, 
  Granada, Spain (rgb@iaa.es). On behalf of \cm.}

\altaffiltext{2}{IAA-CSIC (epm@iaa.es). On behalf of \ac.}

\shortauthor{Garc\'ia-Benito} 
\shorttitle{RevMexAA(SC) Demo Document}

\listofauthors{R. Garc\'a-Benito} 
\indexauthor{Garc\'ia-Benito, R.}

\abstract{\cm\ (COSmos harMONIC) is a sonification project with a triple purpose: 
analysis (by means of sounds) of any type of data, source of inspiration for 
artistic creations, and pedagogical and dissemination purposes. In this contribution 
we present the work recently produced by \cm\ in the latter field, creating specific 
cases for the inclusive astronomy dissemination project \ac\ for blind 
and partially sighted people, but also aimed at a general public that wants to understand 
astrophysics in an alternative format. For this project, \cm\ seeks to create simple 
astronomical cases in their acoustic dimension in order to be easily understood. 
\cm's philosophy for these sonifications can be summarized in a simple metaphor: 
painting graphs with sounds. Sonification is a powerful tool that helps 
to enhance visual information. Therefore, \cm\ accompanies its audios with 
animations, using complementary methods to reach a general public. 
In addition to provide some cases created by \cm\ for inclusive 
astronomy, we also share our experience with different audiences, 
as well as suggest some ideas  for a better use of sonification in 
(global) inclusive outreach.
}

\resumen{(COSmos harMONIC) es un proyecto de sonificaci\'on con un 
triple objetivo: análisis (mediante sonidos) de cualquier tipo de datos, 
fuente de inspiración para creaciones artísticas, y fines pedagógicos 
y de divulgación. En esta contribución se presenta el trabajo realizado 
recientemente por \cm\ en este último campo creando casos específicos 
para el proyecto de divulgación de astronomía inclusiva \ac\ orientado 
no solo a personas con discapacidad visual sino también dirigido a un público 
general que quiera apreciar la astrofísica en un formato alternativo. 
Para este proyecto, \cm\ ha creado casos sencillos en su 
dimensión acústica para que sean fácilmente comprensibles. La filosofía de 
\cm\ a la hora de crear estas sonificaciones se puede resumir con una sencilla 
metáfora: pintar gráficas con sonidos. La sonificación es una poderosa herramienta 
que puede ayudar a reforzar la información visual. Por eso, \cm\ acompaña 
sus audios con animaciones, usando canales complementarios para llegar a una
mayor audiencia. Además de presentar algunos ejemplos creados por \cm\ para 
astronomía inclusiva, se comparte nuestra experiencia con diferentes tipos de público 
así como algunas sugerencias para el futuro de la sonificación en la divulgación 
(global) inclusiva.}

\addkeyword{Sonification}
\addkeyword{Astrophysics}
\addkeyword{Inclusive Outreach}
\addkeyword{Inclusive Cultural Aesthetics}

\begin{document}
\maketitle

\section{What is sonification?}
\label{sec:intro}

Although audification has been around for more than a century as a 
means to listening to data in science, from physiology to seismology
\citep[e.g.,][]{Bernstein1881, Wedenskii1883, Frantti1965},
sonification, as an academic discipline, is a relatively young 
area of research that integrates a wide variety of professional fields. 
As in other media, where digitization produced a radical transformation, 
the 1990s was key to the conceptualization of the phenomenon, in particular, 
with the impulse of the first International Community for Auditory Display 
(ICAD) meeting in 1992.

The general consensus defines sonification as the technique that uses data 
as a source to generate sounds from their transformation, that is, it is a 
representation of data by means of sounds. Sonification would therefore be 
the counterpart of visualization.

Nowadays, sonification is present in our daily lives, from car parking 
assistance to computers. Sounds are used to inform about the success of 
particular events, such as deleting the trash folder (auditory icons), 
email alerts or shutting down the operating system (earcons). They 
can even be the only ingredient to create unique acoustic ecosystems in 
audiogames\footnote{\url{https://www.audiogames.net}}.

Aside from the multiple daily uses of sonification in user-machine 
interactions, there are three main broad areas of applications of 
this technique to scientific data, considering the receptor and objective 
of the output, all of them with wide common overlapping areas: 
a) analysis by means of sounds of any kind of data; 
b) pedagogical and dissemination purposes; 
c) source of inspiration for artistic creations and public 
engagement. Depending on the public and aim of the 
sonification, the acoustic complexity and the clear presence of 
the data will have different weights. 

Composers have been at the fore-front of the artistic use of scientific 
data in their compositions\footnote{For example, 
Lucier's \textit{Music for Solo Performer} premiered in 
1965 \citep{Lucier1967}. For an overview of this 
piece, see \citet{Straebel2014}.}.  
One of the many recent examples is a piece by \citet{Everett2013} 
sonifying chemical evolution using biochemical data exploring 
possible early Earth formations of organic compounds.  
The holistic artistic experience 
\texttt{Chasmata}\footnote{\url{https://www.thedkprojection.com/chasmata/}} 
\citep{Chasmata} with Mars as the leading role, integrates astrogeology, 
architecture, music, sound art, visual art, sculpture, and the venue 
itself, where the audience is part of the creative process using 
their cell phones to interact with more than one hundred musicians 
and multi-channel site-specific ambisonic electronics. The idea of using 
the building as part of the creation has been taken a step further in 
\citet{ExoSounds}, where the piece is performed and recorded  
at the very dome where the sonified data were observed.

However, in general composers are more concerned about the 
final musical experience as a whole rather than the building blocks of their 
creation. Their compositional materials act as a powerful self-inspiring 
creativity engine as well as an exotic component of the story for engaging 
the audience and point their curiosity to the science behind the scenes 
for a later more in-depth reading.

As appealing and aesthetically exciting these works might be, they are not 
easy to use in some specific outreach activities, as the data is embedded in 
rather complex musical setups. There is a broad intermediate land between 
the pure analytical use of sonification and its application for 
compositions. In the following sections we present a couple of cases 
produced by the sonification project \cm\ inhabiting this 
``transitional buffer state''. In section \ref{sec:ref} we will provide some 
reflections on the future of sonification, with special attention 
to outreach. 

\section{The CosMonic Project}

\cm\footnote{\url{http://rgb.iaa.es/cosmonic}} 
(COSmos harMONIC) is a sonification project 
which aims to explore any kind of data 
(passenger rail traffic, weather patterns, 
astrophysical data, $\ldots$) 
in all its acoustic dimensions, from the more analytical 
science-based approach to artistic applications, with particular 
care to the nature of the problem and the audience of the final 
product. In addition, \cm\ is open to collaboration with 
other projects which seek to create sonification 
resources to bring a new and different perspective.

As an example of current collaborations, \cm\ has been working creating 
custom-made cases for the \ac\  
project\footnote{\url{http://astroaccesible.iaa.es/}} \citep{pm17}, 
an inclusive astronomy dissemination project for blind and 
partially sighted people, also aimed at a general public that 
wants to understand astrophysics in an alternative 
format.
For this project, \cm\ seeks to create simple cases in their acoustic 
dimension in order to be easily understood. The idea is to produce brief 
cases that can be used within an inclusive activity to explain or illustrate 
astrophysical concepts. \cm’s philosophy for these sonifications can be 
summarized in a simple metaphor: painting graphs with sounds.

Sonification is a powerful tool that helps to enhance visual content creating 
powerful sinergies thanks to the codification of the information using an 
alternative channel. Therefore, \cm\ accompanies its 
audios with visual animations, combining complementary methods 
so that the described concepts reach the general public. 
In this regard, the audivisual content is inclusive, allowing a broad 
audience to enjoy the information in its multiple dimensions.

All the material produced by \cm\ is uploaded to its own dedicated 
YouTube channel. Each video contain at the beginning a short 
audio description, complemented with a fully detailed explanation 
in the YouTube video description of each sonified case. All the 
input data has been extracted from the literature and/or catalogues 
and credited in the text description. There are two 
versions of each video distributed in two separate YouTube lists: 
one with audio and text descriptions in 
English\footnote{English \cm\ YouTube channel:\\ \yencms} 
and another in 
Spanish\footnote{Spanish \cm\ YouTube channel:\\ \yescms}.

\cm\ does not have a fixed set of tools for producing its products 
(audio or video), as it relies on the best approach for each particular project. 
For the cases described below, a combination of ChucK audio programming 
language \citep{chuck} and Python (video and data pre-processing) 
were used.

\subsection{Case 1: Transits and Soundscapes}
\label{sec:ex1}

The first case is the data sonification of a transit, one of the 
most common techniques used for the hunt for exoplanets. The audio ``paints'' 
the observed data of the transit of the extrasolar planet HAT-P-7b (Kepler-2b), 
located at a distance of 320 pc, discovered in 2008 \citep{hatp7b}. The data was 
collected from the NASA Exoplanet 
Archive\footnote{\url{https://exoplanetarchive.ipac.caltech.edu}}. 
In some cases there are light curves from different observations of the 
same object, both from professional and amateur 
observers\footnote{The data in this case is from amateur observer 
Peter Kalajian (KP2).}. The telescope collects light information from the star at 
regular intervals for several hours (in this particular case). In the visual 
dimension, the horizontal axis represents time (in hours) and the vertical axis 
represents the relative flux. In the audio 
dimension, the brightness value (relative flux) is coded with frequency and the discrete 
nature of the data is represented by assigning a single short sound for each measurement. 
In the animation each individual observations is represented by a blue dot that appears 
in the graphic synchronized with its corresponding sonified sound. The first time an 
observation shows up, to make easier its visual identification, it is plotted in yellow 
for a brief instant. Then, it turns blue and the next observations appear, 
repeating the process until all points are located in the graphic. To facilitate the 
interpretation of the data, a continuous orange line representing the model fit of the 
data is drawn at the same pace as the individual dots.

Since the data spans for an interval of $\sim$ 6 hours, the whole duration of 
the sonification is compressed to 20 seconds. The total duration is a balance 
between the clear individual identification of the ``dots'' and making it easier 
to remember the ``shape'' of the 
observations in the frequency space. An excessively long sonification could make it 
difficult to remember the first part of the data once the audio its reaching its end. 
The relative flux, varying only a few thousandths around 1, is mapped between 
20 and 880 Hz. The scatter of the relative flux (and thus, frequency) is due to 
uncertainties in the measurements. The overall effect is a soft soundscape 
loosely resembling some bits of ``droidspeak''. 


So far we have just ``painted'' the individual quantized data. We could try to extract 
the fit model that represents the overall frequency (relative flux) behaviour of the 
transit, blurring out the individual dots and the intrinsic scatter of the observations. 
This is done in a second sonification using as input the same data. It should be 
noted that we are not sonifiying the data of a model fit curve, but instead we are 
extracting this information directly from the audio data itself. It can be seen as 
a fitting procedure on the audio dimension. This can be done in many ways. We 
chose a fairly simple method based on the use of networks of simple 
allpass and comb delay filters controlled by a mix parameter. The result 
is a dark dense soundscape (``stellar wind'' effect) where the average 
frequency (relative flux) of the dot cloud at a particular instant is enhanced 
over the individual events. 

From this two versions of the same case we can gather two interesting 
ideas. First, we are able to mimic some of the data fitting/visual plotting 
procedures on the audio dimension. And second, we can produce and extract 
very different soundscapes from the same data. We will comment 
some ideas on this in the last section.

\subsection{Case 2: Stellar Evolution and Emotional Reactions}
\label{sec:ex2}

The second case is in fact a series of sonifications on stellar 
evolution, an audio trip through the Hertzsprung-Russell (HR) diagram. 
These cases show the variation of the luminosity and temperature 
of stars along their lifetimes as a function of the initial mass. The data 
was taken from \citet{schaller1992}. Since the goal was to show the 
variation of these two quantities, we fixed the metallicity of the models
to solar (0.02 Z$_{\odot}$) for all the cases. We chose several key 
masses values to illustrate the highly variable evolution of a star as 
a function of its initial mass. Each mass has its one audiovideo: 
1.25 \msun, 5 \msun, 20 \msun, 60 \msun, and 85 \msun. 

In the visual dimension, the logarithm of the temperature (inverted, so 
from left to right the temperature decreases following the usual HR diagram 
convention) is plotted on the horizontal axis, while the logarithm of the 
luminosity goes in the vertical axis. The whole track is already drawn from 
the beginning and a cartoon of a star is placed at the beginning of the track. 
Once the evolution sets off, the star moves along the track. A label 
on the upper left of the plot shows the actual time in units of million years 
as the star moves through its life cycle.

In the audio dimension, luminosity is represented by the frequency (the 
more luminous the star, the higher the frequency) and temperature is 
mimicked by the vibrato, the higher the temperature, the 
higher the rate\footnote{The speed with which the pitch is varied. 
The amplitude of the vibrato is kept constant in these cases.} 
of the vibrato.
This seems a rather intuitive mapping as 
temperature may be related with high activity and vibrations, such 
us boiling water and molecular heat (molecules vibrate faster as 
they heat up). Unlike the first version of the previous case, here 
the sound is continuous and the evolution unfolds in a smooth way. 
The sound of a bell indicates the passage of time, 
struck at regular intervals depending on the lifespan of the star.
All these settings are briefly audio described at the beginning 
of the video. The duration of the sonifications, as in the previous 
case, was set to around 20 seconds.

As some stars, depending on their initial mass, end up exploding as 
supernovae, we have marked this event at the end of the tracks, both 
visually and aurally, by a cartoon and the sound of an explosion.

These cases are good examples of multi-dimensional sonifications, 
where several variables and event informations unfold concurrently: 
from multi-parameter mapping (luminosity-frequency, 
temperature-vibrato) to auditory events in the form of auditory 
icons (supernovae explosion) and earcons (bell). The level 
of complexity in the audio dimension is moderate, but it has 
been tested to work successfully for a general audience. Although 
in general the use of timbre and loudness are the most common 
companions to pitch in multi-parameter mapping, some of 
those acoustic variables (i.e. loudness) do not show an 
effective high dynamical range (in the comfort zone) when used 
simultaneously with others, in particular with pitch. Here the 
use of the vibrato has been proven to perform particularly 
efficient for its discriminating ability.

\section{A few reflections on sonification}
\label{sec:ref}

This last section provides some general reflections 
we believe deserve some further attention, in particular 
(although not entirelly focused on) the realm of outreach, 
dissemination and artistic applications of sonifications.

\subsection{Soundscapes and Emotions}

In section \ref{sec:ex1} we have described two very 
different renditions of the same input data 
information, that create contrasting 
soundscapes, from the more gentle 
``droidspeak'' to the dark dense ``stellar wind''. 
These two distinct sound atmospheres are likely to generate 
diverse reactions and emotions to different audiences. 
The friendly ``droidspeak'' will more likely be preferred 
by young children, while the dark soundscape will be 
the choice for an adult audience, composers or 
sound artists interested in complex acoustic 
environments. This is the case in our experience. 
The ``dark'' soundscape tends to frighten the young  
audience, specially if the illumination of the venue is 
faint (generally, to increase the contrast of the screen). 
On the other hand, this is the preferred option, combined 
with a dark light atmosphere, for more mature audiences 
seeking for deeper experiences. The combination of audio and 
a particular environment setup creates synergies reinforcing 
the emotional side. 

With case \ref{sec:ex2} we have experienced a different 
and more amusing type of reaction in the public. 
The final steps in the evolution of some stars (e.g., listen 
to the 20 \msun\ case) stay during some time in a 
certain region with lower temperature and thus, slow 
vibrato rate. The contrast between high rate vibrato at the 
beginning of the life of the star and the low, maintained 
vibrato rate puts a smile in the public and even makes 
the audience laugh (if presented properly). 

In general, a significant number of sonifications 
of the universe available for outreach
on the internet have a ``spooky'' (audifications in particular) 
spirit or a dark/mysterious character. As exciting as 
these might be, not all ears will have the same (positive) 
reception. We may arouse other types of emotions to 
experience the (otherwise silent) universe.

All these cases illustrate the power of sound 
to evoke quite diverse mood states and point to the 
need to know, where possible, the type of audience 
that will listen to our sonifications, and even 
prepare accordingly the atmosphere of the venue 
to reinforce or dilute some of these effects. 

\subsection{Complex Sonifications and Ear Training}

From high school to college, we have been trained 
for years to interpret increasingly complex plots. 
When we look at a graph, we are not aware of 
this long trained skill that allows us to read 
intricacy relations from the visual point of view. 
All in all, specially in journal articles, we 
encounter dense and elaborated plots that require 
several minutes of our attention. A significant 
fraction of these figures cannot be completely 
deciphered without the help of a detailed caption.

This fact has several implications for sonifications. 
First, we cannot expect a sonification (especially a 
complex one) to be understood without a context and 
description of the content, a function similar to a 
caption figure and a legend. Second, in most cases we 
might need to listen a sonification several times, 
in the same way a plot is visualized for several 
minutes, focusing on different aspects of the data.

The sonification field and its many applications is 
growing and we might adventure it is here to 
stay and become a common source for 
understanding data. 
As any other field, it evolves, creating more 
challenging content, in particular in the science 
and analytical applications. We cannot assume a 
not trained user to handle these complex acoustical 
data as fluently as the visual content. Therefore, 
we argue it would be helpful to prepare future 
sonification users from early stages. Basic 
and advanced ear training from primary to high school 
might be an interesting curricula to implement 
if we aim to deal with rather complex sonifications 
in the future. It should be noted this does 
not imply formal musical 
training\footnote{Needless to say, formal musical 
training would be more than enough for this purpose. 
However, as enriching as this is, not everybody might 
feel attracted to study the complex intricacies of music 
notation or counterpoint rules.}, 
but a more general training focused on 
recovering/interpreting information from 
the aural channel.

\subsection{Inclusive (Cultural) Aesthetics?}

A significant number of outreach/artistic sonifications 
either present mysterious/dark soundscapes or have a 
western classical/pop approach (and instruments). 
Some tools to create these materials rely on technical 
architectures that imprint a cultural bias 
scheme\footnote{For example, generally sounds are 
quantized in twelve fixed chromatic 
notes per octave tuned in equal temperament.} 
and therefore are not particularly suited for 
other musical aesthetics 
\citep{Cornelis2010, Thompson2014}. 

Our strongly rooted cultural and linguistic concepts 
might influence the way we unconsciously picture the 
world. For example, Mandarin speakers in general
conceptualize the future as behind and the past as in 
front of them (also translated into co-speech gestures), 
and those space-time mappings are affected by 
different expressions \citep{Gu2019}. In addition, 
different cultural traditions may have quite different 
instrumental timbres, music scales, aesthetics intervals 
(consonance) or complex polyrhythms that appeal 
in a particular way to the emotions. 
Artistic sonifications can also be 
inclusive from a culturally aesthetic point of view,  
mixing different timbers, styles and even temperaments 
as it has been explored historically in the past \citep{ECDE}.

We have seen that all audiences are not alike and 
can react in many different ways to the same sonic 
source. The (silent) universe does not have a particular 
sound and it should speak in many languages. We should 
encourage sound artists, sound engineers, composers and 
artist in general with experience in or from different 
cultural backgrounds to create a more inclusive aural 
representation of the silent universe.

\acknowledgments{
\footnotesize
R.G.B and E.P.M. acknowledge support from the State Agency 
for Research of the Spanish MCIU through the ``Center of 
Excellence Severo Ochoa'' award to the Instituto de Astrof\'isica 
de Andaluc\'ia (SEV-2017-0709). R.G.B. also acknowledges support 
from grants P18-FRJ-2595 and PID2019-109067-GB100.}


\end{document}